\documentclass[fleqn,usenatbib]{mnras}

\usepackage{newtxtext,newtxmath}

\usepackage[T1]{fontenc}

\DeclareRobustCommand{\VAN}[3]{#2}
\let\VANthebibliography\thebibliography
\def\thebibliography{\DeclareRobustCommand{\VAN}[3]{##3}\VANthebibliography}

\usepackage{graphicx}	
\usepackage{amsmath}	
\usepackage[normalem]{ulem}
\usepackage[dvipsnames]{xcolor}
\usepackage{hyperref}
\usepackage{academicons}
\usepackage{scalerel}
\usepackage{tikz}
\usetikzlibrary{svg.path}

\newcommand{\sfr}{M$_{\odot}$ yr$^{-1}$}

\newcommand{\ucre}{u$_{\rm{CRE}}$}
\newcommand{\ugas}{u$_{\rm{gas}}$}
\newcommand{\umag}{u$_{\rm{mag}}$}

\newcommand{\kmps}{km s$^{-1}$}
\newcommand{\msun}{M$_{\odot}$}
\newcommand{\ngc}{NGC $4631$}

\definecolor{orcidlogocol}{HTML}{A6CE39}
\tikzset{
  orcidlogo/.pic={
    \fill[orcidlogocol] svg{M256,128c0,70.7-57.3,128-128,128C57.3,256,0,198.7,0,128C0,57.3,57.3,0,128,0C198.7,0,256,57.3,256,128z};
    \fill[white] svg{M86.3,186.2H70.9V79.1h15.4v48.4V186.2z}
                 svg{M108.9,79.1h41.6c39.6,0,57,28.3,57,53.6c0,27.5-21.5,53.6-56.8,53.6h-41.8V79.1z M124.3,172.4h24.5c34.9,0,42.9-26.5,42.9-39.7c0-21.5-13.7-39.7-43.7-39.7h-23.7V172.4z}
                 svg{M88.7,56.8c0,5.5-4.5,10.1-10.1,10.1c-5.6,0-10.1-4.6-10.1-10.1c0-5.6,4.5-10.1,10.1-10.1C84.2,46.7,88.7,51.3,88.7,56.8z};
  }
}

\newcommand\orcid[1]{\href{https://orcid.org/#1}{\mbox{\scalerel*{
\begin{tikzpicture}[yscale=-1,transform shape]
\pic{orcidlogo};
\end{tikzpicture}
}{|}}}}

\title[Radio Halo of NGC 4631]{Radio Halo of NGC 4631: Comparing Observations and Simulations
}

\author[Vijayan et al]{
Aditi Vijayan\orcid{0000-0002-7714-2379}$^{1,2,3}$\thanks{E-mail: aditiv@rri.res.in},
K . S. Dwarakanath$^{1}$,
Biman B. Nath\orcid{0000-0003-1922-9406}$^{1}$,
Ruta Kale\orcid{0000-0003-1449-3718}$^{4}$
\\
$^{1}$Raman Research Institute, Bangalore, 500080, India \\
$^{2}$Shanghai Astronomical Observatory, Chinese Academy of Sciences, 80 Nandan Road, Shanghai 200030, China\\
$^{3}$Inter-University Centre for Astronomy and Astrophysics, Post Bag 4, Ganeshkhind, Pune 411 007, India\\
$^{4}$National Centre for Radio Astrophysics, Tata Institute of Fundamental Research, S. P. Pune
University Campus, Ganeshkhind, Pune 411007, India
}

\date{Accepted XXX. Received YYY; in original form ZZZ}

\pubyear{2021}

\begin{document}
\label{firstpage}
\pagerange{\pageref{firstpage}--\pageref{lastpage}}
\maketitle

\begin{abstract}

We present low frequency observations at $315$ and $745$ MHz from the upgraded Giant Metrewave Radio Telescope (uGMRT) of the edge-on, nearby galaxy \ngc. We compare the observed surface brightness profiles along the minor axis of the galaxy with those obtained from hydrodynamical simulations of galactic outflows. These are 3D simulations that replicate star-formation in a Milky-Way mass galaxy and follow magnetized outflows emerging from the disk. We detect a plateau-like feature in the observed emission at a height of $2-3$ kpc from the mid-plane of the galaxy, in qualitative agreement with that expected from simulations. This feature is believed to be due to the compression of magnetic fields behind the outer shocks of galactic outflows. We model the observed surface brightness profiles by assuming an exponential as well as a Gaussian fitting model. Using $\chi^2$ statistics, we find that the exponential model fits the profiles better and we use it to determine the scale heights. We estimate the scale height for the synchrotron radio emission to be $\sim 1$ kpc. The timescales for advection due to outflows and diffusion of cosmic ray electrons are $\gtrsim 5$ and $\sim 160$ Myr, respectively. Because advection acts on a timescale much shorter than diffusion, we conclude that in \ngc\ advection, rather than diffusion, plays the dominant role in the formation of radio halo. The spectral index image with regions of flatter radio spectral index in the halo appears to indicate possible effects of gas outflow from the plane of the galaxy.

\end{abstract}




\section{Introduction}
Outflows related to star formation processes, such as supernovae and stellar winds, can expel gas from the disk of a galaxy to many kilo parsecs above it. Such outflows are multiphase in nature and thus, can be studied over a wide range in the electromagentic spectrum. Multi-wavelength observations of this extra-planar gas reveal the details of physical mechanisms responsible for its ejection from the disc. While X-ray studies give clues to the physical state of gas at temperatures $\gtrsim 10^6$ K \citep{Strickland&Heckman07, Strickland&Heckman09}, optical lines probe the gas at intermediate temperatures ($\sim 10^5$ K) \citep[see for e.g.]{Heckman+15} and molecular lines give information about cooler gas ($\lesssim 10^4$ K) \citep{Walter+02}. 

Non-thermal interactions taking place in the outflows can be studied via observations in the radio band. Supernovae can create strong shocks, which are sites of cosmic ray electron (CRE) acceleration. Synchrotron radiation is produced when CREs interact with the magnetic field. Extended synchrotron emission around a galaxy is called ``radio halo'' \citep[see for e.g.]{Dahlem+95, Dahlem+06, Irwin+12}. Properties of such halos around nearby star forming galaxies, such as their extent and shape, have been qualitatively linked to star formation processes occurring in the disk.

In order to quantify these links, morphological and spectral observations need to be compared with numerical simulations. We have previously performed hydrodynamic (HD) and magneto hydrodynamic (MHD) simulations of star forming galaxies in order to understand extra-planar thermal and non-thermal emission from star-forming galaxies \citep{Vijayan+18, Vijayan+20b}. There are several ways of incorporating the effects of cosmic ray (CR) propagation into simulations, such treating CR as a fluid \citep{Gupta+21} or a collection of particles \citep{Mignone+20}. In this paper, we discuss the results from \cite{Vijayan+20} (referred to as V20 in the rest of the paper), which estimates CR energy density by employing different assumptions about how CRE gain energy- via total gas energy, gas thermal and kinetic energy or magnetic energy density. It is further assumed that each of these modes lead to different equipartition relations for the CRE energy density. Using energy considerations, it was concluded that equipartition between the CRE energy density and the total gas energy density produced results most consistent with observations \citep{Dahlem+95}. The simulations in V20 also predict the existence of a ``plateau'' in the one-dimensional surface brightness profiles of synchrotron emission along the minor axis of the galaxy. Both the predictions from simulations, viz, equipartition relation of CRE energy density and the presence of a plateau region- can be tested using observations. In this respect, we choose the nearby MW-mass galaxy, \ngc, to test these predictions. We chose \ngc because earlier observations appear to indicate the existence of such a plateau based on the Ooty Radio 
Telescope observations at $327$ MHz \citep{Sukumar&Velusamy85}.

In the simulations of V20, CRs propagate via advection through galactic outflows and such halos are referred to as ``advection-dominated halos''. Advection-dominated halos form when winds generated from SNe feedback carry CRs to high altitudes and can be accompanied by steepening of the spectral index \citep{Irwin+99}. Observations have found that if CREs are able to diffuse out of the disks of the galaxies, ``diffusion dominated'' radio halos can also be formed \citep{Dahlem+95, Irwin+99}. Advection and diffusion occur on different timescales depending on independent physical parameters such as wind outflow speed and diffusion constant, respectively. Previous observations have distinguished between advection and diffusion dominated halos by comparing these timescales with the synchrotron lifetime of CREs \citep{Heesen+09, Krause+18}. For high outflow speeds, i.e., smaller advection timescales, CR diffusion may not play a dynamically important role in the early evolutionary stages of galactic outflows \citep{Jana+20}.

In this paper, we re-examine the synchrotron radio emission from \ngc\ using the new high resolution radio images from the upgraded Giant Metrewave Radio Telescope (uGMRT). In Section \ref{sec:theoretical_motivation}, we discuss the theoretical motivation for and main results from V20. Section \ref{sec:ngc4631} summarises the motivation for choosing \ngc. In Section \ref{sec:observations}, we discuss the low frequency observations of \ngc\ carried out using the uGMRT. In Section \ref{sec:results}, we compare the observations data with the simulations carried out by us earlier. We conclude the paper by discussion in Section \ref{sec:discussion} and present the main conclusions in Section \ref{sec:conclusions}.

\section{Theoretical motivations}\label{sec:theoretical_motivation}

We briefly summarise the simulation results related to the radio halo of a star forming galaxy presented in \cite{Vijayan+20}, which provide the theoretical framework for comparison with observations of \ngc. We consider a Milky Way-sized galaxy, with a warm ($\sim 10^4$K) rotating disk, surrounded by a hot ($\sim 10 ^6$ K) halo. The initial density and pressure distribution of the gas within the simulation domain is dictated by the dark matter potential of a $10^{12}$ \msun\ halo. At the beginning of the simulation, there is hydro-dynamic equilibrium between the disk and the halo.
The simulation domain is a box with spatial extent [$x$, $y$, $z$]= [$\pm 6$] kpc and the disk is taken to be lying in the $x-y$ plane. Feedback from supernova occurs through $1000$ identical OB associations, which are distributed across the volume of the disk, following the Kennicutt-Schmidt law for distribution along the radial axis and a gaussian distribution along the vertical axis (see Appendix A of \cite{Vijayan+18} for details). The total star formation rate is divided equally over these $1000$ injection points.
The warm disk is magnetised in the azimuthal, $\hat{\phi}$-direction such that the magnetic field has an exponential profile in the radial and vertical directions,  $\hat{R}$) and $\hat{z}$, respectively.
Quantitatively, the initial magnetic field has only an azimuthal component, and is given by (adapted from \cite{Sun+08}),
\begin{equation}\label{eqn:bfield}
B_\phi=B_0 \exp \Bigl [-{\frac{R}{R_0}} -\frac{\vert z \vert}{z_0}\Bigr ] \,, 
\end{equation}
where $B_0$ is $2$ $\mu$G and $z_0$ and $R_0$ are equal to $400$ pc and $1.5$ kpc, respectively. This value was chosen to ensure stable equilibrium for the initial setup, in which the study of the effect of stellar feedback can be carried out.

In order to estimate the synchrotron emissivity from the simulations, we also need information about Cosmic Ray electron (CREs) energy density (\ucre). For the analysis, it was assumed that cosmic rays constitutes a fraction $0.1$ of the total gas energy density, \ugas, which includes contribution from thermal and kinetic pressure. Of this CREs account for a fraction of $5\%$ of  the total CR energy budget. Therefore, \ucre\ is equal to a fraction, $\epsilon_{\rm{CRE}}$ ($=0.5$\%), of \ugas. Then, with the assumption of a power-law energy spectrum for CR electrons ($n(E) \propto E^{-p}$), with an index $p=2.2$, we calculated the emissivity in each grid point and produced a map of the radio halo at a given frequency. We direct the reader to V20 for further details related to the setup and the results.

\begin{figure}
	\includegraphics[width=\columnwidth]{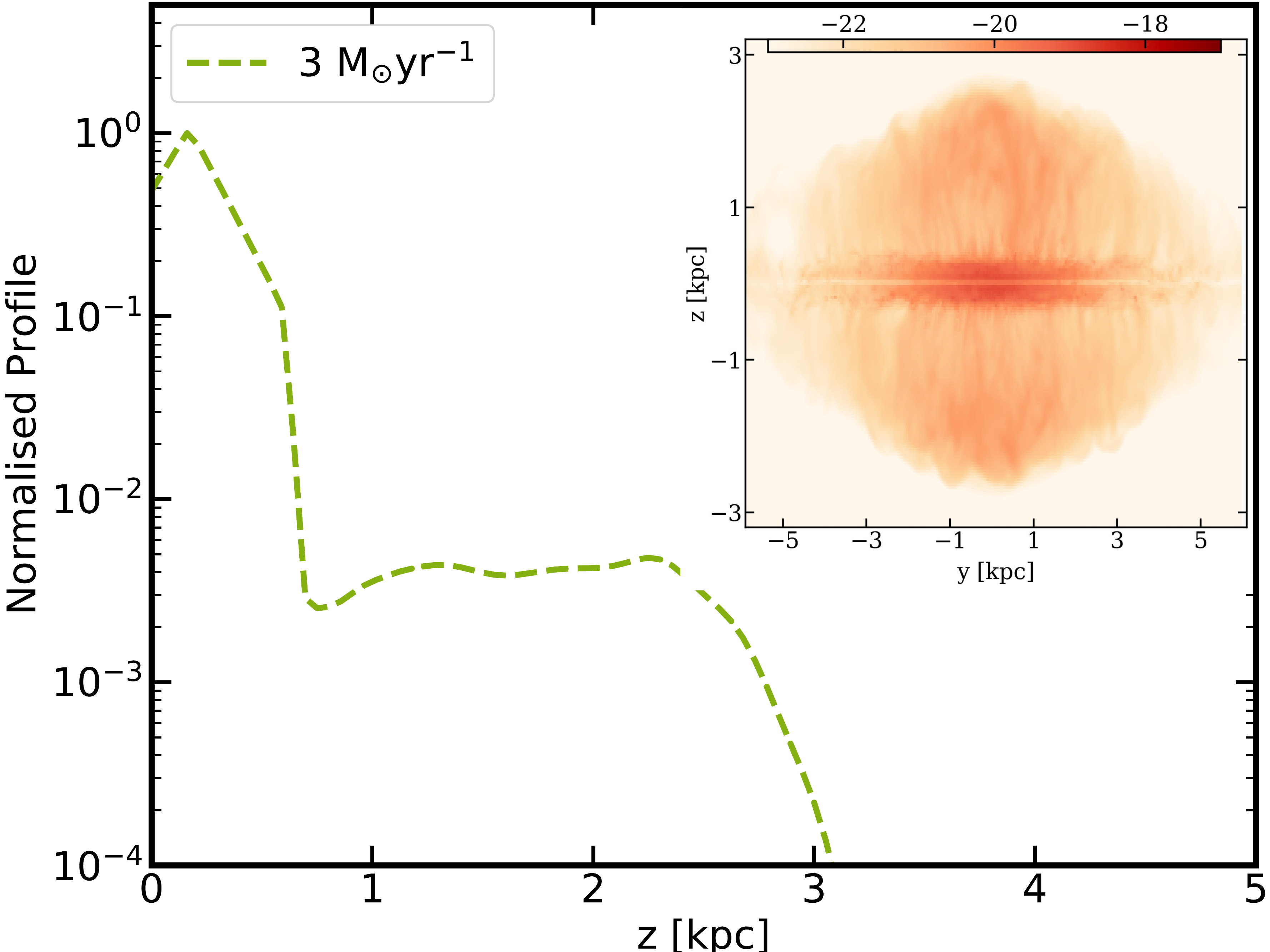}
    \caption{One dimensional surface brightness profile along the minor axis of the 
galaxy at $1.4$ GHz for a star formation rate of $3$ \sfr. 
In the inset, we show the simulated surface brightness image at the 
same frequency after a $5$ Myr evolution. The colour bar indicates the brightness in units of log[erg s$^{-1}$ cm$^{-2}$ Hz$^{-1}$sr$^{-1}$]. }
    \label{fig:line-brightness-sim}
\end{figure}

Coherent energy injection through the OB associations produces large scale galactic outflows, which emerge from the disk within $\sim 1$ Myr of evolution. The outflows are magnetised as they carry the magnetised disk gas to regions outside of the disk. In the inset of Figure \ref{fig:line-brightness-sim}, we show the two-dimensional surface brightness maps made using the simulation data, after $5$ Myr of evolution, from the run with a SFR of $3$ \sfr, which corresponds closely to the SFR estimates of NGC $4631$ (see Section \ref{subsec:sources_of_radio_emission} for a discussion). We produce these 2D surface brightness maps by integrating the synchrotron emissivity along the line of sight, taken to be $x-$ axis. 
We note that the radio emission is strongest for the disk region because it contains the strongest magnetic field as well as gas density. The brightness decreases for extra-planar regions as the gas rarefies. However, the gas collects behind the shock as it moves into the ambient medium. The accumulation of magnetised matter produces enhanced emission from these regions. 

This enhanced emission is more prominent in the one dimensional surface brightness profiles (referred to as ``surface brightness profiles'' in subsequent text) along the minor axis of the galaxy. We generate surface brightness profiles by summing the 2D surface brightness along the $y-$axis and plotting it along the $z-$axis, which is the minor axis of the simulated galaxy. We show these profiles for the same SFR in Figure \ref{fig:line-brightness-sim}. The simulated surface brightness drops sharply from the disk region ($\lesssim 500$ pc) as the outflows expand and then, plateau within the outer shock ($\sim 2-3 $ kpc) region. Near the position of the outer shock, $\sim 2\hbox{--}3$ kpc, we notice a factor of few increase in the surface brightness corresponding to its outer boundary.

\section{NGC 4631}\label{sec:ngc4631}

The plateau in the surface brightness profile in our simulations, produced by the accumulation of magnetised plasma uplifted from the disk, motivated us to look for observational signatures of the same.

In order to observe the plateau, as seen in Figure \ref{fig:line-brightness-sim}, over a distance, $|z| \sim 1 - 2$ kpc, we require a source with a bright emission in the radio band. The emission in the radio band correlates with the SFR, thus we would like to observe a star-forming galaxy. Further, we need a close-by galaxy to ensure that the spatial resolution in the vertical direction is adequate to resolve the extent of the plateau. \ngc\ is currently undergoing tidal interaction with its companion, NGC $4656$.

\ngc\ is a near-by ($\sim 7$ Mpc), edge-on galaxy with a moderate star formation rate equal to $\sim 3$ \sfr \citep{Strickland+04a}. It has been observed at a number of frequencies, viz. $327$ MHz (OSRT observations, \cite{Sukumar&Velusamy85}), $610$ MHz, $1412$ MHz (WSRT observations, \cite{Ekers+77}) and $10.7$ GHz \citep{Klein+84}. At $610$ MHz this galaxy has a radio extent of $15' \times 10'$. These observations were made with relatively poor resolutions by present day standards ($\sim 1^\prime$ at $327$ \& $610$ and $0.5^\prime$ at $1412$ MHz). A plateau-like feature in the surface brightness profile is indicated in the observations of \cite{Sukumar&Velusamy85} at $327$ MHz. 

Most recently, \ngc\ was observed as part of the EVLA survey \citep{Irwin+12}. These observations were carried out at $1.5$ GHz with the VLA-C array. As the authors have already noted, the most extended features of NGC $4631$ were not recovered in these observations. The extent of the radio halo was measured to be $1.8-2.3$ kpc \citep{Krause+18, Mora-Partiarroyo+19}. \cite{Mora+19} undertook a detailed analysis of the orientation and magnitude of the magnetic field in the disk as well as the halo region of \ngc. In the eastern part of the disk of \ngc\, the magnetic field orientation is parallel to the major axis, which makes the comparison with the simulations of \cite{Vijayan+20} reasonable (see Section \ref{sec:comparison_with_sims} for details). Despite existence of many observations, no detailed analysis of the NGC $4631$ halo (nor a high resolution imaging) has been carried out. 

There are circumstantial evidences for an accumulated layer of gas in other nearby starburst galaxies. The vertical intensity profile in CO emission for M82 shows a hint of a plateau at $\sim 1$ kpc \citep{Leroy+15}. However, radio continuum emission from this region is likely to be missed in high resolution images because of its extension and low surface brightness. Therefore a closer galaxy with high SFR, and preferably an edge-on galaxy would be most suitable for this experiment, and NGC $4631$ fits the requirements. Further, there are evidences of expanding molecular shells near the centre of \ngc as traced by $^{12}$CO(J$=1-0$) \citep{Rand2000} and $^{12}$CO(J$=3-2$) spectral lines \citep{Irwin+11}. 

\section{Observations}\label{sec:observations}
The galaxy \ngc\ was observed with the Giant Metrewave Radio Telescope (GMRT) in the wideband system during $20-28$ January, 2020. A continuous track of $8$ hours was carried out in each of Band $3$ ($250-500$ MHz) and Band $4$ ($550-900$ MHz). The basic data was recorded with an integration of time of 8s and 2048 channels. The calibrators used were 3C286 and 3C147 and tied to the Perley-Butler flux density scale of 2017 \citep{Perley&Butler2017}. The data was analysed using the fully automated pipeline for making images from the upgraded Giant Metrewave Radio Telescope (uGMRT) called CAsa Pipeline-cum-Toolkit for Upgraded Giant Metrewave Radio Telescope data REduction - CAPTURE. It is a python program that uses tasks from the NRAO Common Astronomy Software Applications (CASA) to perform the steps of flagging bad data, calibration, imaging and self-calibration \citep{Kale&Ishwara-Chandra21}. The salient features of the pipeline are a fully automatic mode to reduce the raw data to produce a self-calibrated continuum image, specialized flagging strategies for short and long baselines that ensure minimal loss of extended structure, and flagging of persistent narrow band radio frequency interference (RFI). The data were subjected to deconvolution and self-calibration loops with $4$ phase-only and $4$ amplitude and phase cycles. The deconvolution was carried out using the Multi-Scale, Multi-Term, Multi-Frequency Synthesis mode with W-Projection which is a well-established deconvolution and imaging method for synthesis arrays like the Very Large Array and the GMRT \citep{Cornwell+08, Rau&Cornwell2011, Rau+16}. Three scales corresponding to $1$, $5$ and $15$ times the synthesized beam were used in the deconvolution. A value of $2$ was used for the Taylor terms in the task CLEAN to account for the curvature in the spectra across the bandwidth of the observations. Larger number of scales and/or higher Taylor terms did not make any further improvement to the images. 

Images of different resolutions and sensitivities were produced by varying the weighting of the visibilities (robust parameter in the task CLEAN). Varying the robust parameter \citep{Briggs1995} from $-2$ to $+2$ changes the weighting of the visibilities from `uniform' to `natural'. Natural weighting produces images sensitive to extended emission, but with a loss of resolution. Uniform weighting, on the other hand, produces images with the best resolution, but at the loss of sensitivity to extended emission. The final primary-beam corrected images adopted here are the ones produced with robust $= 0$, which is a compromise between sensitivity to extended emission and resolution. Based on the known largest extent of this source from single dish (Ooty Radio Telescope) \citep{Sukumar&Velusamy85} measurements ($\le\sim$ $15^{'}$) and the analysis of recovery of extended emission from uGMRT observations \citep{Deo&Kale2017} it is clear that $\ge\sim$ 90$\%$ of the emission from this galaxy is recovered in these observations.
t

\section{Results}\label{sec:results}
\begin{figure}
	\includegraphics[width=\columnwidth]{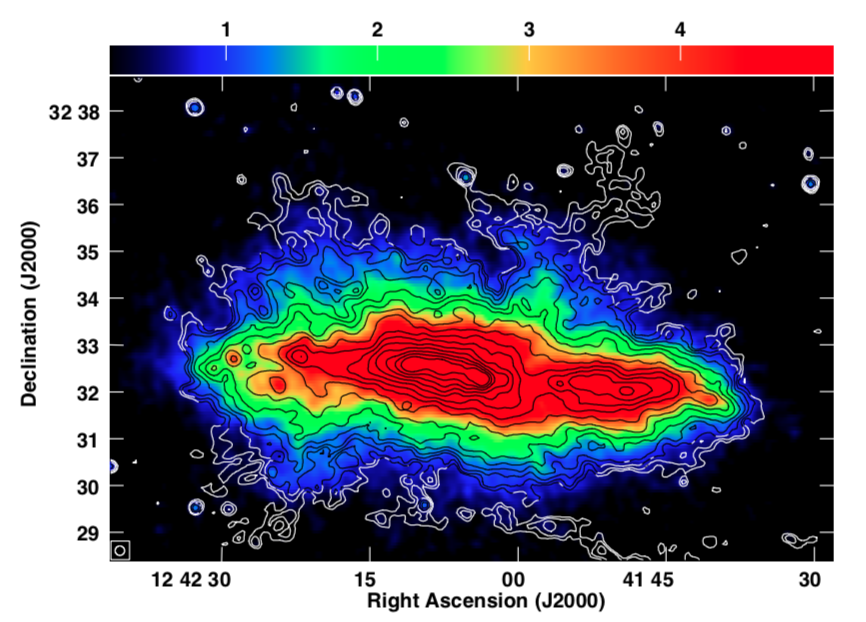}
    \caption{Radio image at $745$ MHz (contours) from uGMRT overlaid on the image at $315$ MHz (color). Both the images have a resolution of 12$^{''}$ as indicated at the lower left corner. The $745$ and $315$ MHz images have RMS ($\sigma$) values of $50$ and $80$ $\mu$Jy ~beam$^{-1}$ respectively. The first contour is at 3$\sigma$ and increases in steps of $\sqrt 2$. The color scale ranges from $0.24$ to $5$ mJy ~beam$^{-1}$.}
    \label{fig:radio-2freq}
\end{figure}

Figure \ref{fig:radio-2freq} shows the images at a resolution of $12^{''}$, from Bands $3$ ($\nu_{\rm center} = 315$ MHz) and $4$ ($\nu_{\rm center} = 745$ MHz). 
These images indicate that \ngc\ has an extended synchrotron emission.
We note that the emission is strongest in the disk region of the galaxy, as expected. In the $745$ MHz images, we also note the presence of filamentary structures rising from the disk region. From Figure \ref{fig:radio-2freq}, we note that this galaxy is 
tilted with respect to the image axis. 
We correct for the position angle of the major axis and align it horizontal.

\begin{figure*}

	\includegraphics[width=\textwidth]{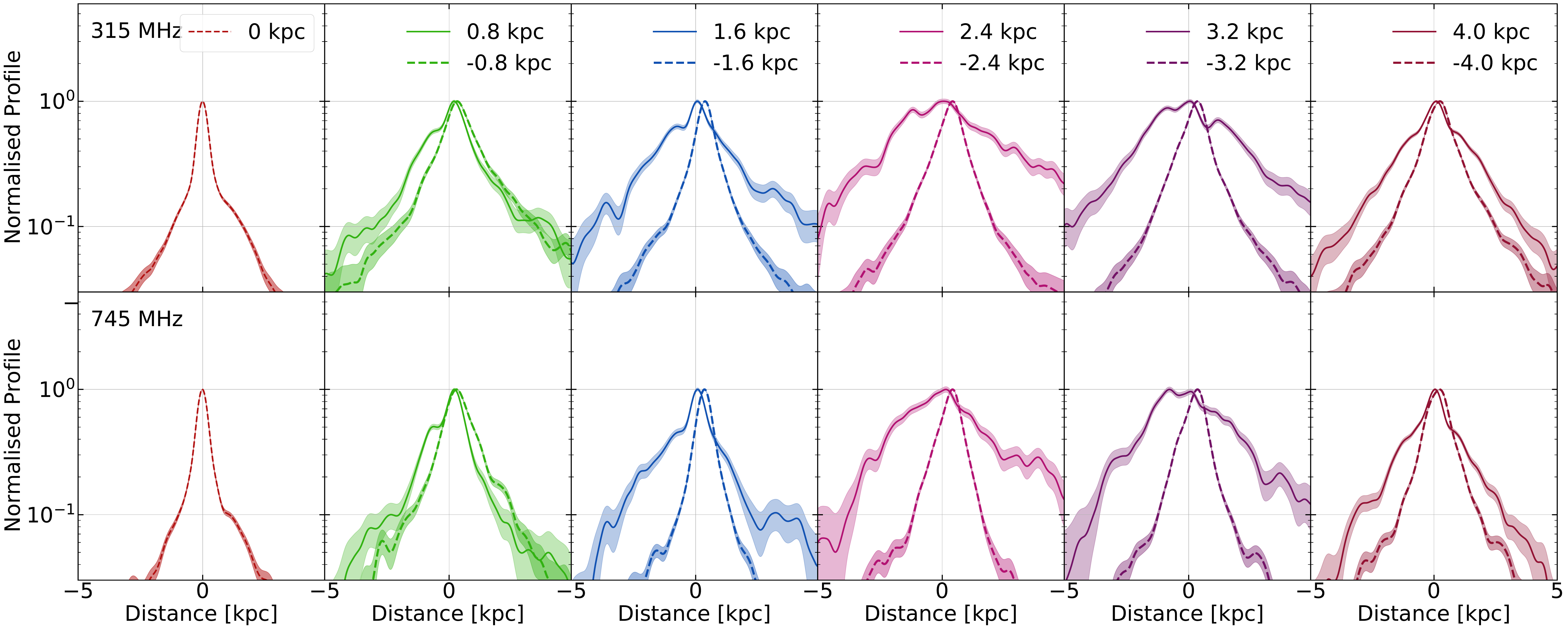}
    \caption{Brightness profiles along the minor axis at $315$ MHz (top) and $745$ MHz (bottom). The values at the top right hand corner (in the top panel) indicate the distance from 
the center along the major axis at which the profiles were extracted. 
The value of 0 kpc corresponds to $\alpha=$ 12h42m3.35s, 
$\delta=$ +32$^{0}32^{'}16.5^{''}$ and is approximately the middle of the major axis. 
The positive (negative) values at the top right corner 
correspond to the West (East)
of the $0$ kpc respectively.
The profiles were taken 24$''$ apart (2 times the resolution) 
on the major axis which corresponds to $0.8$ kpc. The shaded regions indicate $\pm3\sigma$ error bars.}
    \label{fig:line-profile-315-745}
\end{figure*}

\begin{figure}
	\includegraphics[width=\columnwidth]{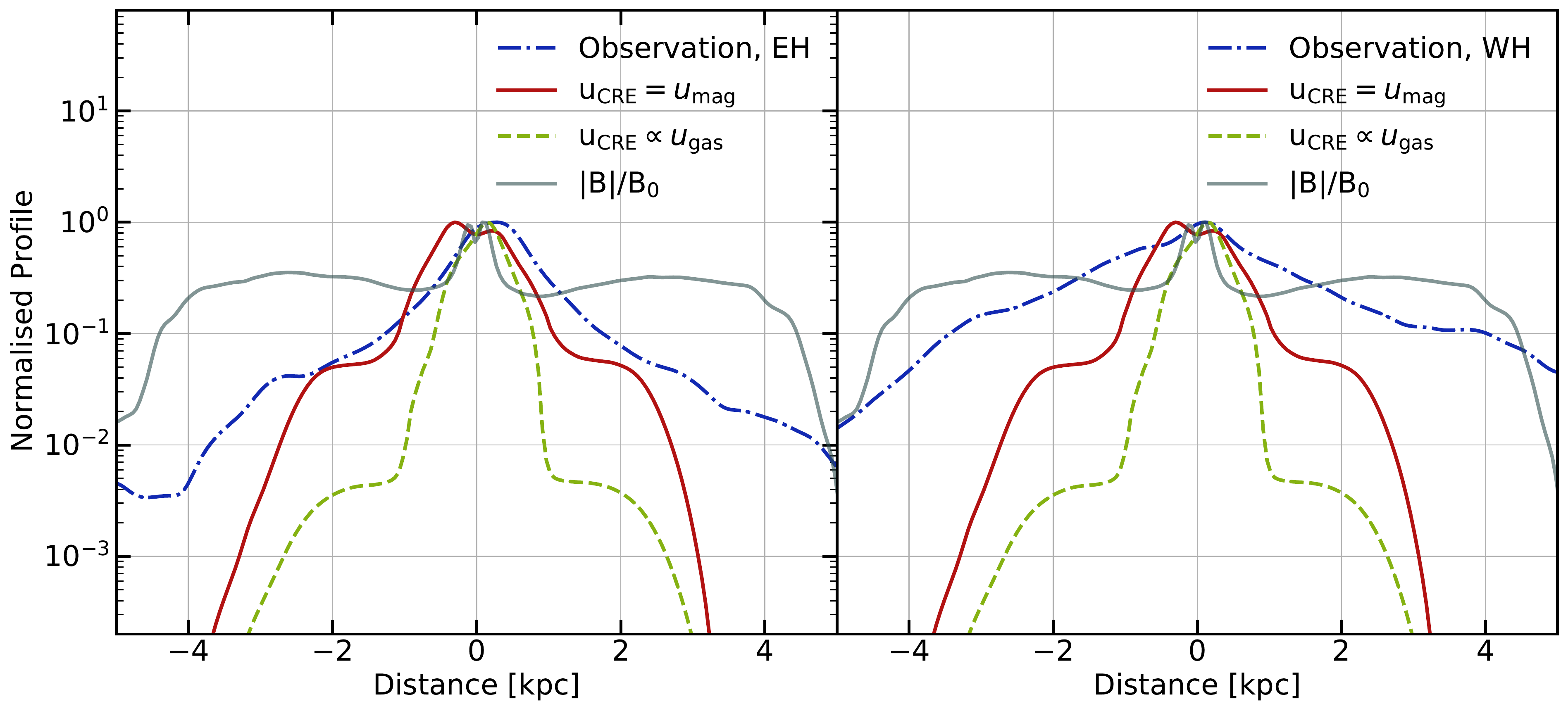}
    \caption{Comparison of observed brightness profiles at $\nu=745$ MHz with the 
simulated profiles.
The solid red curve shows the case of equipartition between the cosmic ray and 
magnetic field energy densities, while the dashed curve refers to the case when the cosmic ray energy density is a fraction of the gas density as discussed in the text. 
The dot-dashed curve shows the observed profile. The observed profile in the left
panel was obtained after  averaging over all the dashed curves in 
Figure \ref{fig:line-profile-315-745} corresponding to 745 MHz. 
Similarly, the observed profile in the right panel 
was obtained by averaging all the solid curves in Figure \ref{fig:line-profile-315-745}
corresponding to $745$ MHz.
Gray curves represent the strength of the magnetic field in the simulation 
after $5$ Myr of evolution, scaled by the midplane value. }
    \label{fig:ucre-umag-comparison}
\end{figure}

\subsection{Comparison with Simulations}\label{sec:comparison_with_sims}

In Figure \ref{fig:line-profile-315-745} we show the observed surface brightness profiles at the two frequencies, normalised by their values at the mid-plane. Each panel in the Figure represents the surface brightness profile along the minor axis of the galaxy at a given distance along the midplane of the galaxy. The positions in the midplane, with respect to the centre of the galaxy, at which these vertical profiles were extracted are shown in each panel as the distance along the major axis from the origin. We limit our analysis to $\pm 4$ kpc of the centre of the galaxy to ensure a fair comparison with the simulations, for which the extent of star formation is $\pm 4$ kpc. 
These profiles are plotted against the vertical distance along the minor axis of the galaxy (taken to be the $z$-axis). 

The profiles at both the frequencies show a monotonic decline as a function of distance from the midplane. While many of the profile show a plateau-like feature, the profiles at $745$ MHz shows plateau-like feature prominently at a distance of $z\sim 2.5$ kpc from the mid-plane. These features are similar to the ones seen in the simulated profiles. While the intensity of the simulated profile in Figure \ref{fig:line-brightness-sim} decreases by $\sim 4$ orders of magnitude within $\sim 1$ kpc from the mid-plane, the observed profile follows a much shallower decline as it decreases by only $\sim 1.5$ orders from the mid-plane value.

To compare the simulated and the observed profiles more quantitatively, in Figure \ref{fig:ucre-umag-comparison} we show the observed and simulated surface brightness profiles at $745$ MHz, averaged along the major axis. Differences between the simulated and observed profiles can arise due to the way \ucre is estimated. If \ucre\ is assumed to be a fraction of the \ugas\ (Section \ref{sec:theoretical_motivation}), the simulated surface brightness profiles depart hugely from the observed profiles (Figure \ref{fig:ucre-umag-comparison}). Figure \ref{fig:ucre-umag-comparison} also shows the simulated profile if \ucre is taken to be in equipartition with the magnetic field energy density (solid red curve) \footnote{Neglecting spectral ageing.}. 
In the outflows, the gas and magnetic energy densities are comparable (see Figure $3$ of \cite{Vijayan+20}). Therefore CREs are more energetic when we take \ucre\ to be equal to \umag. The profiles produced with \ucre\ in equipartition with \umag\ are broader and lead to stronger synchrotron emission. The ratio between the peak and the brightness at $\sim 2$ kpc height is of order $\le 10$ in the observed and simulated profiles in the case of the equipartition of CRE with the magnetic field. \cite{Mora-Partiarroyo+19} also suggest that in the radio halo of \ngc\, equipartition is established between CR and magnetic field as the latter couples to the outflowing gas.

Despite the relative increase in the intensity of the simulated profiles
by changing the prescription for \ucre, we note that the 
observed profile is shallower than the simulated profiles by a factor of a few.
Such differences can arise because of the lower value of the assumed 
magnetic field in the simulations. We know that the synchrotron intensity has a steep magnetic field dependence, $\propto B^{3+\alpha}$ \citep{Beck&Krause05}. 
In Figure \ref{fig:ucre-umag-comparison}, we also show the 
profile of the magnetic field strength relative to its 
value at the midplane after $5$ Myr of evolution, 
the time step at which we extract the simulated surface brightness profiles. 
Although the vertical magnetic field is exponential initially, 
stellar feedback processes modify it over time. 
At the epoch when these profiles are examined, the 
relative magnetic field strength profile drops in the vicinity of the disc, but 
quickly becomes shallower, dropping by a factor of $\sim 3$ at a 
height $\approx 2$ kpc from the mid-plane. 
This is consistent with the findings of \cite{Mora2013}, who 
reported a decrease in the field strength by a factor of $1.5$ 
from the mid-plane to the halo region (from $13\pm2\,\mu$G to $10\pm2\,\mu$G). In fact, if the magnetic field at $\sim 2$ kpc increases by $\sim 20\%$, the intensity will increase by a factor of $2$, which is adequate to explain the discrepancy between the observed and simulated profiles. 

The discrepancy between the simulated and the observed profiles can also be a result of the tidal interactions between \ngc\ and its companion NGC $4656$, which is at a distance (in projection) of $\sim 60$ kpc. An examination of the total HI content images \citep{Rand1994} reveals that there are a number of tidal features emerging out of the disk of \ngc\ due to the tidal interactions. A comparison of the radio images at $745$ MHz (Figure \ref{fig:radio-2freq}) with the total HI content images reveals that many of the continuum emission plumes out of the disk closely follow the tidal strips in HI. The tidal interactions likely account for the observed differences between the simulated and observed brightness profiles. Furthermore, the tidal interactions are also likely to be responsible for the different brightness profiles observed along the Eastern and Western parts of the \ngc\ disk.

\subsection{Sources of radio emission}\label{subsec:sources_of_radio_emission}

Based on the uGMRT observations, the integrated flux density of \ngc\ at $315$ MHz is $3.6$ Jy at $315$ MHz. For an assumed distance of $6.7$ Mpc (see Table 1 of \cite{Radburn-smith2011}) and a spectral index of $-0.8$, the implied SFR is $3.4$ M$_\odot$ yr$^{-1}$ using the relationship in \cite{Yun2001}. This value is close to the SFR value ($\sim 3$ M$_\odot$ yr$^{-1}$) inferred by \cite{Strickland+04} from {\it IRAS} flux density.

If we use the H$\alpha$ flux of $14.3 \times 10^{40}$ erg s$^{-1}$ \citep{Hoopes+99}, and the correlation found by \cite{Kewley2002}, we obtain a SFR of $\sim 1.1$ \sfr. We note that \cite{Smith+01} determined $E(B-V)\approx 0.33$ for the galaxy, which corresponds to a visual extinction of roughly $A_v\approx 1$. In other words, the dust extinction in the visible band is significant for NGC $4631$, which probably explains the low H-$\alpha$ flux and consequently the low value of inferred SFR from it.

The H$\alpha$ luminosity allows us to estimate the free-free contribution to the radio continuum power at $315$ MHz \citep{Lequeux1980}. We find that free-free contribution is 
an order of magnitude lower than the total observed radio flux density. Therefore, even if the true H-$\alpha$ flux is a few times larger than observed, the main contribution to the radio continuum at $315$ MHz would still be from synchrotron emission.

\subsection{Spectral Index}\label{sec:spec_index}
\begin{figure}
	\includegraphics[width=\columnwidth]{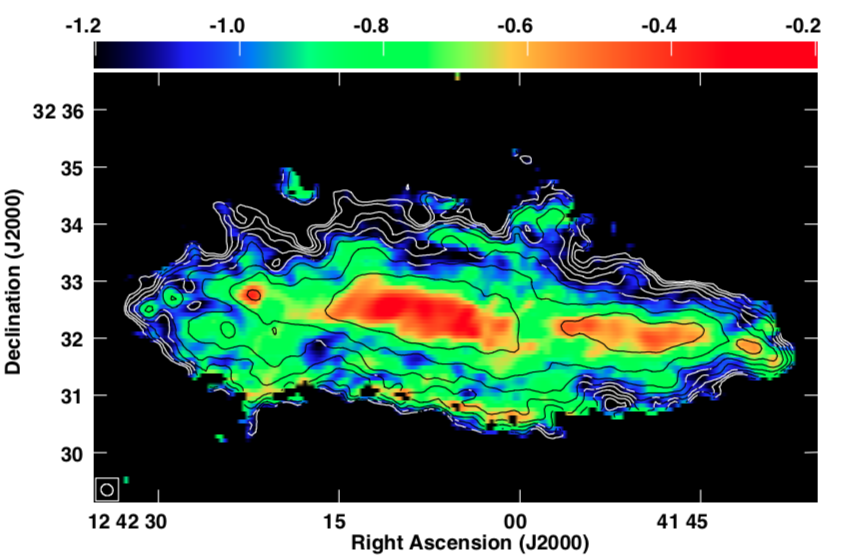}
    \caption{Spectral Index Image (color) obtained between $315$ and $745$ MHz images from uGMRT. The color scale in spectral index ($\alpha$) ranges from $-0.2$ to $-1.2$, where, S $\propto \nu^{\alpha}$ and S is the flux density. The innermost contour starts at 0.025 and increases in steps of 0.025 and indicates the error on the value of the spectral index.}
    \label{fig:spec-index}
\end{figure}

\begin{figure}
	\includegraphics[width=\columnwidth]{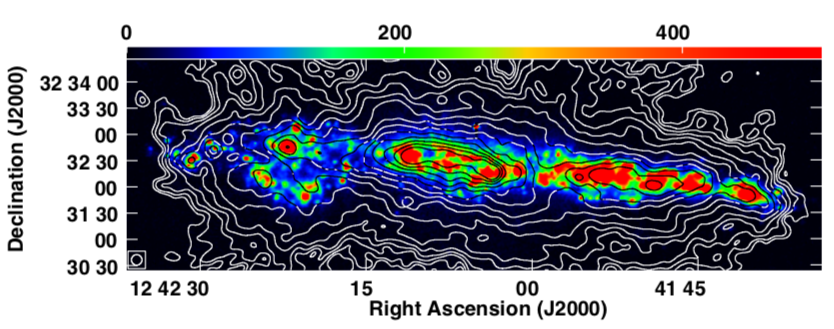}
    \caption{Radio image at $745$ MHz (contours) 
 overlaid on H$\alpha$ image (color). Contour levels are the same as in Figure \ref{fig:radio-2freq}.}
    \label{fig:halpha-radio}
\end{figure}

Figure \ref{fig:spec-index} shows the spectral index image produced using the images from Bands $3$ and $4$. Given the wide bandwidth of the uGMRT observations
($\sim 400$ MHz), the visibility coverage at both the frequencies is excellent and
continuous all the way down to $\sim 100 \lambda$. Such a distribution of visibilities
makes it straight forward to estimate the spectral index distribution across this source
whose maximum extent is $\sim 15'$.
The red patches in the image around $\delta \approx 32.5^{\circ}$ correspond 
to regions of flatter spectrum ($\alpha >0.6$). These regions are also bright in the H$\alpha$ image of the 
galaxy (see Figure \ref{fig:halpha-radio}), indicating that these correspond to star formation sites.
The radio emission from these regions is likely a mix of synchrotron and free-free emission.

In the extraplanar regions, the spectral index become progressively steeper with distance from mid-plane on average.  Such a steepening is expected to take place due to synchrotron energy loss of electrons over a timescale determined by the cooling timescale (for emission at $31$5 MHz), $t_{\rm loss} \approx 80 B_{10\, \mu \rm{G}}^{-3/2}$ Myr, where $B=10 \, \mu{\rm G} \times B_{10 \, \mu \, \rm {G}}$. We note that \citet{Mora2013} have inferred a mean magnetic field strength of $9\pm 2\mu$G, averaged over the entire disc of NGC $4631$ from depolarization observations (their Table 6), $13\pm2\,\mu$G in the central part of the disc and $10\pm2\, \mu$G in the radio halo region. \citet{Mora+19} find that the magnetic field in the halo of \ngc is coherent with a strength of $\approx 4$ $\mu$G.

\begin{figure*}

	\includegraphics[width=0.8\textwidth]{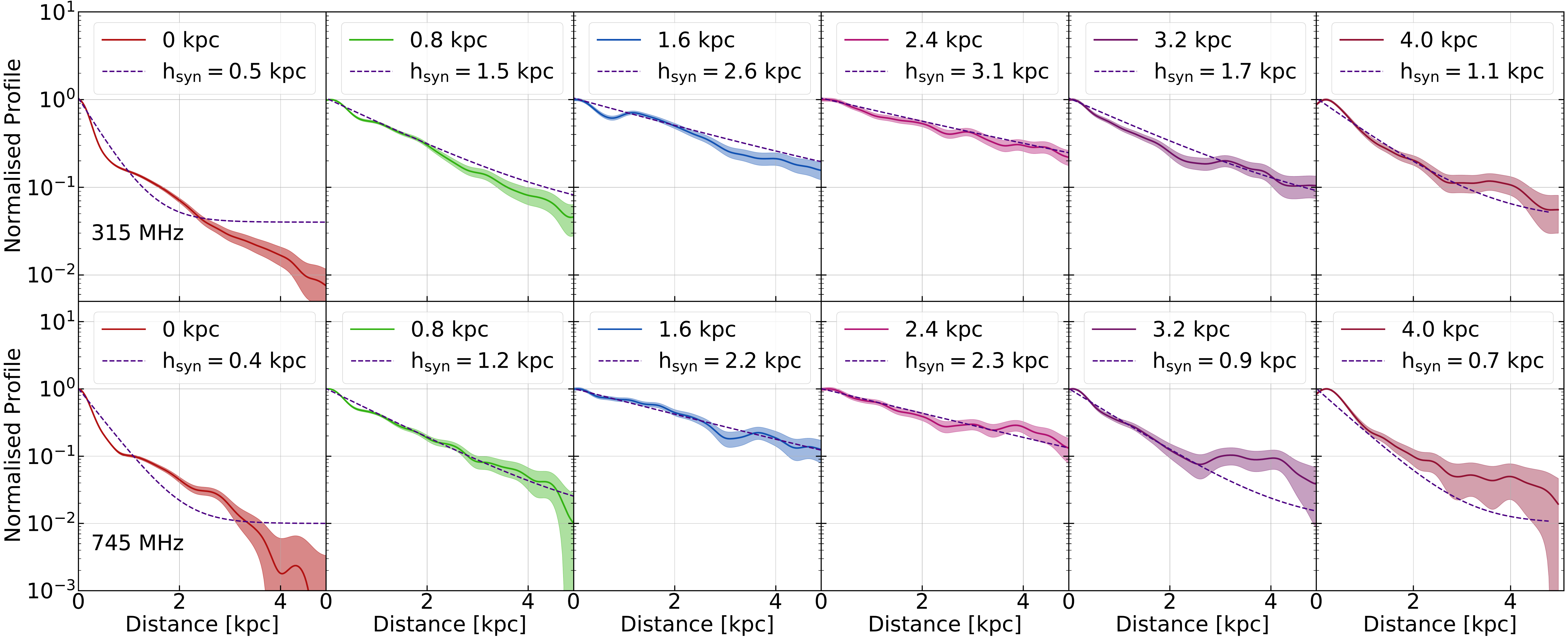}
    \caption{Best fit for the exponential model (given by Equation \ref{eqn:scale_height_exp}) to brightness profiles shown in Figure \ref{fig:line-profile-315-745}. The profiles above the midplane (along the positive values of $x-$axis in Figure \ref{fig:line-profile-315-745}), extracted at positive distances, are shown in solid lines, while the best-fits from Equation \ref{eqn:scale_height_exp} are shown in dashed lines. The scale height, using which the profiles are fit, are indicated in each panel. We produce similar fits for the observed profiles below the midplane (see Figure \ref{fig:scale_height_average}). We note that the fits are rather poor for $|z|\gtrsim 2$ kpc. As in Figure \ref{fig:line-profile-315-745}, the shaded regions indicate $\pm3\sigma$ error bars.}
    \label{fig:scale_height_fit}
\end{figure*}

The cooling timescale in the extra-planar regions can be used to gain insights on transport phenomenon, viz advection and diffusion, which cause electrons to move away from the galaxy mid-plane. Advection refers to the process by which electrons from the disk region are carried to the extra-planar regions by the action of outflows. The advection timescale, $t_{\rm adv}$, occurs over a period determined by the outflow speed and is given by $t_{\rm adv}\approx 10 \, z_{\rm kpc} \upsilon_{100}^{-1}$ Myr. Here, $z_{\rm kpc}$ is the distance reached in units of $1$ kpc and $\upsilon_{100}^{-1}$ is the outflow speed in units of $100$ km s$^{-1}$. We can estimate $\upsilon_{\rm adv}$ from the correlation between $\upsilon_{\rm adv}$ and surface density of star formation, $\Sigma_{\rm SFR}$, that is $\upsilon_{\rm adv} \propto \Sigma_{\rm SFR}^{0.3}$ \citep{Heesen+18, Vijayan+20}. We use the surface density of star formation, $\Sigma_{\rm SFR} \sim 0.06 \rm{M}_{\odot} \rm{yr}^{-1} \rm{kpc}^{-2}$, assuming SFR $\sim 3$ \msun yr$^{-1}$ over radial extent of $4$ kpc. This gives $\upsilon_{\rm adv} \sim 600$ \kmps. Note that although the magnetized plasma accumulates at $\sim 2.5$ kpc in our simulation run for $5$ Myr, the outer shock is at $\sim 3$ kpc, which reflects a bulk speed of $\sim 600$ \kmps, comparable to the  estimate from the above-mentioned scaling.
The timescale for advection is then $\sim 2\, z_{\rm kpc}$ Myr. For vertical distances of a few kpc, the advection timescale is much shorter than the cooling timescale, indicating that the CREs are advected out of the galaxy before they are able to cool.

CREs can also diffuse out of the disk region and into the extra-planar regions. The timescale for diffusion is $t_{\rm diff} \simeq 30\, z_{\rm kpc}^2/D_{28}$ Myr, where $D_{28}$ is the diffusion constant in units of $10^{28}$ cm$^2$ s$^{-1}$.

The diffusion timescale for CREs is, therefore, 
longer than the advection timescale, for $z_{\rm kpc}\le 0.06 \, {\rm kpc} \, D_{28}$. This timescale becomes comparable to the CRE energy loss timescale at
$\sim 1.6 \, D_{28}^{1/2}$ kpc. However, the flux at this height is too small for the spectral index to be measured. But the above estimates indicate that 
advection, and not diffusion, is the main process by which CREs are lifted to the radio halo of NGC $4631$.

\subsection{Scale Height}
\begin{figure}
	\includegraphics[width=0.8\columnwidth]{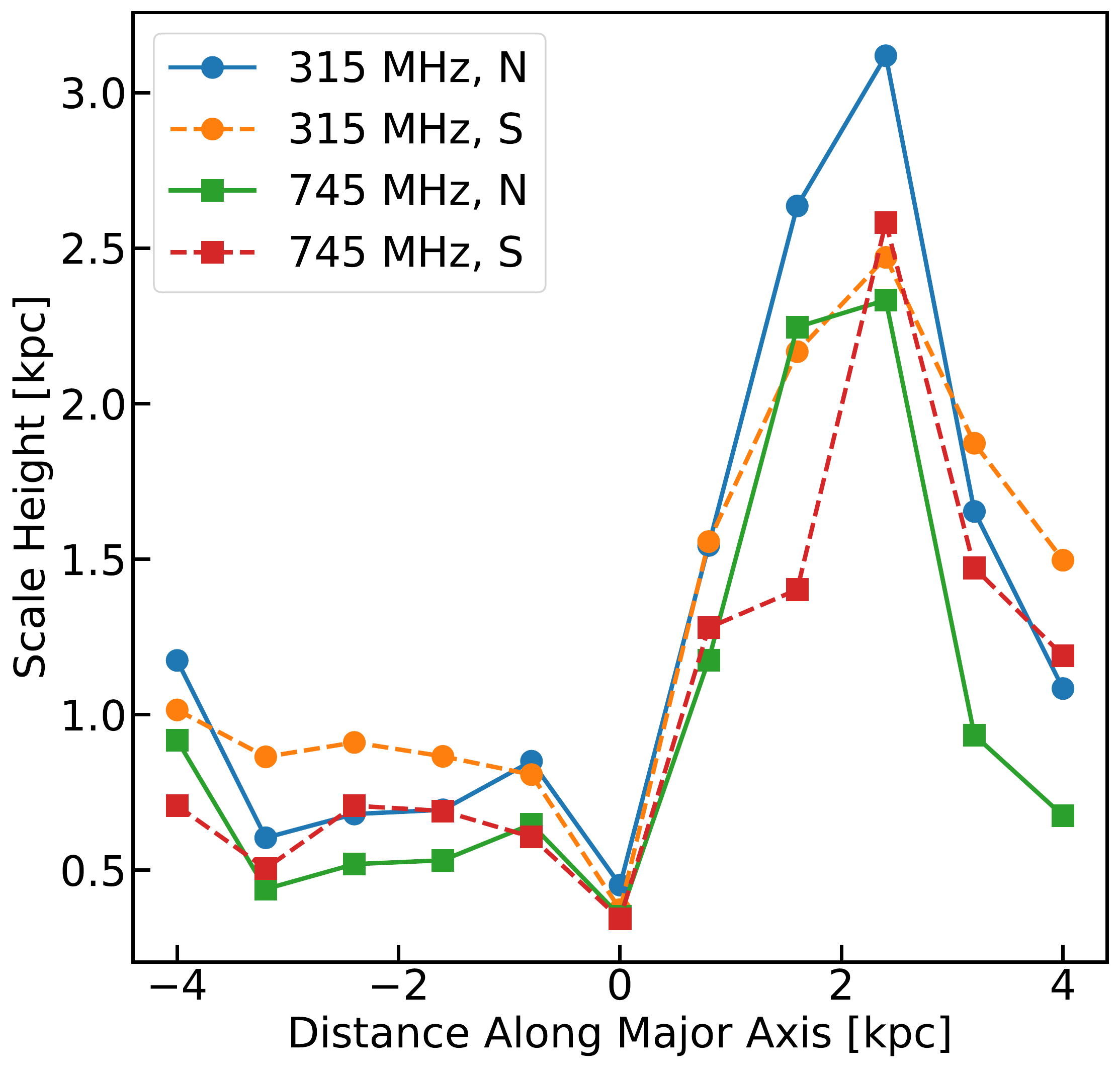}
	\caption{The estimated halo scale heights at different distances along the major axis of the galaxy. The labels 'N' and 'S' refer to the halo scale heights above and below the mid-plane respectively. The $x-$axis represents the distance from the center along the major axis from East (negative) to West (positive).}
    \label{fig:scale_height_average}
\end{figure}

We established in the previous Section that the synchrotron halo of NGC $4631$ is an advection dominant halo. We explore the characteristics of an 
advection dominated halo by estimating the synchrotron scale heights of radio emission of the galaxy. For an 
advection halo $t_{\rm adv} \sim t_{\rm loss}$. Because $t_{\rm loss} \propto B^{-3/2} \nu^{-1/2}$ and $ t_{\rm adv} \propto \upsilon_{\rm adv} t_{\rm adv}$, we can write \citep{Krause+18}
\begin{equation}
    l_{\rm adv} \propto B^{-3/2} \nu^{-1/2}\,.
\end{equation}

This implies that at the observation frequencies of $315$ and $745$ MHz, the $l_{\rm adv}$ are related in the following manner,
\begin{equation}\label{eqn:ldiff_correlation}
    l_{\rm adv,315M} = 1.5 \times l_{\rm adv,745M} \,,
\end{equation}
where $ l_{\rm adv,315M}$ and  $l_{\rm adv,745M}$ refer to the diffusion length scales at $315$ and $745$ MHz, respectively. Further for an advection dominated halo, $l_{\rm adv}$ can be taken to be close to the synchrotron scale height, $h_{\rm syn}$. Strictly speaking, Equation \ref{eqn:ldiff_correlation} is valid only when the synchrotron loss time is comparable to the dominant escape time (here, advection). But as this is not the case for \ngc, we find that the data do not follow the relation.

We use the formulations described in \cite{Dumke+95} to find the synchrotron emission scale height from the observation data. 
Following \cite{Dumke+95} and \cite{Krause+18}, we assume that the observational profiles can be represented by either an exponential profile, given by, 

\begin{equation} \label{eqn:scale_height_exp}
    I_{\nu} = I_{\rm 0,exp} \exp\Big({\frac{-|z|}{h_{\rm syn,e}}}\Big) + I_{{\rm halo},\nu},
\end{equation}

or by a Gaussian profile, given by, 

\begin{equation} \label{eqn:scale_height_gauss}
    I_{\nu} = I_{\rm 0,gauss} \exp\Big({\frac{-z^2}{h_{\rm syn,g}^2}}\Big) + I_{{\rm halo},\nu}
\end{equation}

where $h_{\rm syn,e/g}$ is the synchrotron scale height corresponding to either fits. We find that adding a constant, $I_{{\rm halo},\nu}$ to these
models provides a better fit to the observed surface brightness profiles. We chose the value of $I_{{\rm halo},\nu}$, normalised by the peak, as $2.5\times 10^{-3}$ and $0.9\times 10^{-3}$ for $315$ and $745$ MHz.

We fit the profiles shown in Figure \ref{fig:line-profile-315-745} to 
Equations \ref{eqn:scale_height_exp} and \ref{eqn:scale_height_gauss} independently and obtain the best fit values of $h_{\rm syn,e}$ and $h_{\rm syn,g}$, respectively. Profiles above and below the midplane, referred to as the north (`N') and south (`S') parts, respectively, are also fit independently. 
To establish which of the two profiles are a better fit, we estimate the reduced $\chi^2 (=$(Observation-Fit)$^2$/$\sigma^2$, where $\sigma$ is the standard deviation in the observed data) between the observed profiles and Equations \ref{eqn:scale_height_exp} and \ref{eqn:scale_height_gauss}. The lower the value of the reduced $\chi^2$ for a particular model, the better is the fit. By comparing the respective $\chi^2$ values, we find that the exponential fits the observed profiles better than the Gaussian. The scale heights derived by using the exponential fits are referred to as `$h_{\rm syn}$' in the rest of the paper.

In Figure \ref{fig:scale_height_fit}, we show the best fit exponential profiles (dashed lines) 
along with the observed brightness profiles (solid lines
along the positive distance in \ref{fig:line-profile-315-745}).
We note here that the fits are rather poor for $|z|\gtrsim 2$ kpc at large distances from the centre of the galaxy.
This is likely due to the tidal interactions of \ngc\ with its neighbour as mentioned earlier.

The best-fit synchrotron scale heights, $h_{\rm syn}$, are indicated in the top 
right hand corners in each panel (Figure \ref{fig:scale_height_fit}). The scale height of synchrotron emission is not constant along the disk. To understand how it varies along the midplane, in Figure \ref{fig:scale_height_average}, we show 
$h_{\rm syn,e}$ at different distances along the major of the axis of the galaxy. We note that for the positive distances along the major axis (``western'' part' of the halo), scale height increases and then decreases away from the midplane, while it remains nearly constant in the eastern part. 
This is also seen in Figure \ref{fig:line-profile-315-745} in which the surface brightness profiles extracted at positive distances of the major 
axis are broader compared to those extracted at negative distances. 
The observed synchrotron scale height differences between the Eastern and the Western
parts of the disk are likely due to the tidal interactions that \ngc\ has 
undergone. Variation of synchrotron emission scale heights along the major axis of the 
galaxy have indeed been observed previously \citep{Heesen+18}.

The radio halo is asymmetric about the midplane. The average scale height in the north (south) part is 
$1.32$ ($1.31$) and $0.98$ ($1.04$) kpc for $315$ and $745$ MHz, respectively. These asymmetries could be in part due to the ongoing tidal interactions of \ngc\ with its companion. These scale heights are smaller than the observed at higher frequencies \citep{Krause+18, Mora-Partiarroyo+19}. We find that the scale heights do not exactly follow the correlation expressed in Equation \ref{eqn:ldiff_correlation} as the advection lengths at $315$ MHz are underestimated by $\sim 20\%$ for both north and south parts. This could be because the magnetic field suffers a nominal decrease in its value over the length scales given by the two advection lengths. We note here that scale heights alone are insufficient evidence to discriminate between advection and diffusion dominant halos as in case of escape of CRE the ratio of scale heights can become unity.

In the case of equipartition between magnetic and particle energy density, the magnetic field scales with synchrotron intensity as $B \propto I^{1/(3+\alpha)}$. We also know that the equipartition implies $n_e \propto B^2$. Then, with the vertical profile of $I \propto \exp (-z/h_{\rm sym})$, the vertical profile of the CR electron number density is given by,
\begin{equation}
    n_e \propto \Bigl [ \exp \Bigl ( {-z \over h_{\rm syn}} \Bigr )\Big]^{{2 \over 3 + \alpha}} \,,
\end{equation}
which implies a CR electron scale height of 
\begin{equation}
    h_e={3 + \alpha \over 2} h_{\rm syn} \,.
\end{equation}
Given $\alpha\sim 0.8$, we can infer $h_e \approx 2.3$ kpc, for $315$ MHz. 
Using this value of $h_e$, and the scaling derived in Section \ref{sec:spec_index}, the advection time scale is $\sim 5$ Myr and the diffusion time scale is $\sim 160$ Myr.


\section{Discussion}\label{sec:discussion}


The spectral index distribution across \ngc\ (Figure \ref{fig:spec-index}) has some interesting features. 
The mid-plane emission has a spectral index in the range $-0.4$ to $-0.6$. The spectrum of
the radiation in this region estimated here is somewhat
flatter by $\sim 0.15$ compared to the spectral index distribution in this region estimated
between $1.57$ and $5.99$ GHz using the Effelsberg and the VLA observations (\cite{Mora+19}). 
A detailed look at all the observations and the data concerned lead to the interesting 
conclusion that the spectrum of synchrotron radiation is actually flatter in the 
lower frequency range ($< 1$ GHz) compared to that in the higher frequency range ($>$ 1 GHz). 
Such a flattening of the galactic synchrotron emission spectrum towards lower frequencies
in the mid-plane has been reported for the Milky Way, M51 and NGC 891 \citep{Platania+98, Oliveira+08, Orlando+08, Mulcahy+14, Mulcahy+18, Kierdorf+20}. A more
detailed discussion of the spectral flattening is beyond the scope of the current study. 
The presence of large amounts of ionized gas in the mid-plane of NGC4631 as seen in the 
H$\alpha$ emission from this region (Figure \ref{fig:halpha-radio}) could also be playing
a role in the flattening of the spectrum in the mid-plane.
Outside the mid-plane, the spectral index reaches a value of $-0.8$ to $-1.0$, similar to 
the values reported by \cite{Mora+19} in these regions. 
 Overall, the spectral index monotonically increases to $-1.2$ towards the outer regions of the galaxy. However, there are small regions which show departures from this monotonic trend and show a flattening of the spectral index. This is particularly visible near the southern edge of the galaxy, where, the spectral index flattens to $-0.6$ (orange coloured patches) after having steepened to a value of $\sim -1.0$ in the inner regions.

\begin{figure}
	\includegraphics[width=\columnwidth]{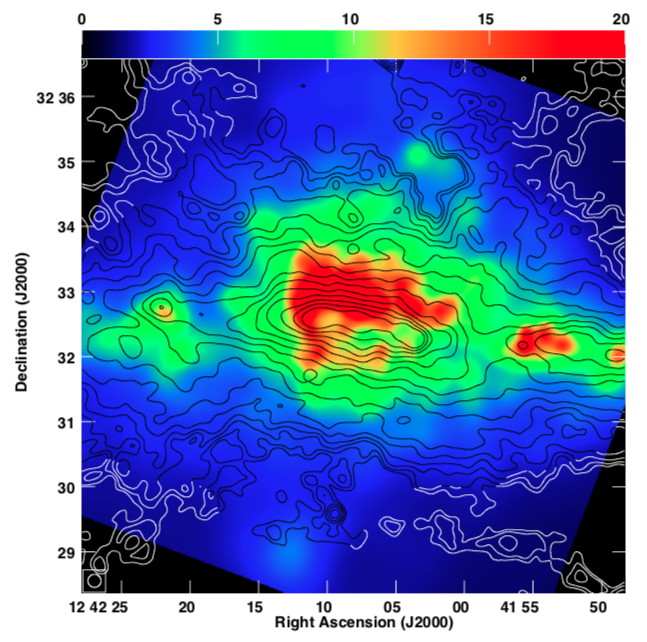}
    \caption{Radio image at 745 MHz (contours) 
 overlaid on X-ray image (color). Contour levels are the same as in Figure \ref{fig:radio-2freq}.}
    \label{fig:xray-radio}
\end{figure}

The flatter spectral index regions seen towards the southern edge of the galaxy (Figure \ref{fig:spec-index}) could be indicative of current shock acceleration at these sites. It is possible that the X-ray images of this galaxy could throw some light on this issue. In Figure \ref{fig:xray-radio} the X-ray image from \textit{Chandra} is overlaid on the uGMRT image at $745$ MHz \citep{Wang+95, Wang+01}. The morphology of X-ray emission is asymmetric with respect to the mid-plane of the galaxy. There is extended X-ray emission towards the northern part of the galaxy with much less of such an extended emission towards the southern parts. Given the intense level of star formation occurring in the mid-plane of the galaxy and the morphology of X-ray emission it appears reasonable that a significant part of the X-ray emission is caused by the outflow of hot gas due to stellar activities and explosions. In such a scenario, if the outflow encounters larger amount of matter towards the southern side, it is expected to stall and create shock waves. The regions of flatter radio spectral index seen towards the southern edge of the galaxy could be a result of such a phenomenon. Such a density difference between the northern and southern part of the mid-plane can also explain the observed X-ray morphology. 
We note that such a possibility has also been invoked in the case of the Milky Way in order to explain the asymmetry of the Fermi Bubble \citep{Sarkar+15} on the two sides of the disc. Alternatively, the star clusters in \ngc\ may be situated above the midplane thus increasing the possibility of creating chimneys and breakthrough of the hot gas on the northern side. 

We note that the simulations run for a period of $5$ Myr, which is much shorter than the cooling or diffusion timescale of CR. The reason is that the simulations were focused on the magneto-hydrodynamical aspect of outflows. The inferences on the CR and radio halo depended on the assumptions made regarding equipartition of magnetic field and CRs. Our conclusions are therefore circumscribed by these limitations, and the robustness of the conclusions can be tested in future with longer simulation runs, including physical effects such as CR diffusion.

We also note that the advection speed as inferred from the comparison with simulation may be an overestimate. \cite{Heesen+18} estimated from spectral index modelling an advection speed of $\sim 200$ \kmps and a scale height (at low frequencies) of $\sim 4$ kpc. This gives an advection time scale of $\sim 20$ Myr, a few times longer than that estimated by us. One might be inclined to infer that the shorter time scale would imply that NGC 4631 is non-calorimetric and the radio brightness should be smaller than that predicted by the star formation rate, which is not the case. However, the shorter advection time scale does not necessarily mean a non-calorimetric case for \ngc, because the accumulation of magnetized plasma above the disk may work towards confining CRs and making NGC 4631 nearly calorimetric.

\section{Conclusions}\label{sec:conclusions}

We report the observations of the edge-on galaxy \ngc\ at $315$ and $745$ MHz with uGMRT, which allow us to study the effect of star formation on the extraplanar gas, and to do a comparison with our previous hydrodynamical simulations. The focus was to study the possible accumulation of radio plasma above a certain height from the disc and its manifestation in radio continuum images. The main results can be summarized as follows- 
\begin{enumerate}

\item We detect a plateau-like feature in the vertical radio surface brightness profiles at a height of $2-3$ kpc from the mid-plane of the galaxy. This is more pronounced in the $745$ MHz images of the galaxy. The plateau-like feature is consistent with the simulations of a star forming galaxy, in which the magnetic field and CR particles are carried out of the disk by gaseous outflows generated by stellar feedback.

\item A quantitative comparison with the simulated profiles implies that equipartition between cosmic ray energy density and magnetic field energy density is at play. 

\item The spectral index image appears to indicate possible effects of shocks produced from outflows of hot gas from the mid-plane.

\item The observed profiles follow the exponential model (Equation \ref{eqn:scale_height_exp}) and we find that the scale height of the radio halo of \ngc\ is $\sim 1$ kpc at both  $315$ and $745$ MHz. 

\item  A comparison of the respective timescales and the scale heights at different frequencies reveals that the advection is dominant over diffusion in carrying the CR particles to the scale height and for forming the radio halo. The advection timescale is a few Myr, whereas the diffusion (for particles responsible for radio emission in observed wavelengths) timescale is an order of magnitude higher than this.

\end{enumerate} 

\section*{Acknowledgements}

We thank the staff of the GMRT that made these observations possible. GMRT is run by the National Centre for Radio Astrophysics of the Tata Institute of Fundamental Research. We would also like to thank Daniel Wang, Richard Rand, and Silvia Carolina-Mora for providing the X-ray, H$\alpha$ maps and high frequency radio data respectively. We thank Rainer Beck and Volker Heesen for several comments and discussions. RK acknowledges the support of the Department of Atomic Energy, Government of India, under project no. $12-$R\&D-TFR-$5.02-0700$. 

\section*{Data availability}
The data underlying this article will be shared on reasonable request to the corresponding author.



\bibliographystyle{mnras}
\bibliography{references} 







\bsp	
\label{lastpage}
\end{document}